# Design of a Quantum Source of High-Frequency Gravitational Waves (HFGW) and Test Methodology.

Giorgio Fontana

*Department of Information and Communication Technology, University of Trento, 38050 POVO (TN), Italy*

*Phone: +390461883906, e-mail giorgio.fontana@unitn.it*

**Abstract.** The generation of High-Frequency Gravitational Waves (HFGW) has been identified as the required breakthrough that will lead to new forms of space propulsion. Many techniques have been devised to generate HFGW, but most of them exhibit marginal efficiency, therefore the power emitted in form of gravitational waves (GW) is orders of magnitude lower than the input power. The gravitational wave counterpart of the LASER, termed Gravitational-wave LASER or "GASER" is the quantum approach to the efficient generation of gravitational waves. Electrons, protons, muons, etc, all have charge and mass, if accelerated they usually lose energy through the very fast electric and magnetic channels, this causes a negligible emission through the gravitational channel. Quantum systems can be engineered to forbid electric and magnetic transitions, therefore the gravitational spin-2 transitions can take place. A class of active materials, suitable for making a GASER based on electronic transitions in the solid state, is identified along with their relevant physical properties. Means for creating coherence and population inversion and means to increase the emission probability are described. The expected performances of the device are derived from quantum gravitational theories. Additional properties of the active materials are considered to enforce the theoretical foundation of the device. A proof-of-concept device, operating at about 1 THz, is described. Experiments are proposed as a natural starting point of the research.

## INTRODUCTION

Present space propulsion technology, mainly based on rocketry, suffers of serious limitations if the range of operations is extended above earth orbit.

Dynamical and speed limitations are accompanied by safety concerns. Rockets must transport the payload and the propellant used as reaction mass. Rockets require such a large amount of energy to accomplish their tasks that management of the required fuel may expose the passengers to a high risk. The energy requirement for a space transportation system can be reduced if a new technology is developed to employ distant masses as reaction mass. The propellant will be no longer needed and will not take away energy.

### Gravitational Wave Propulsion

A remarkable advancement in space propulsion is a form of gravitational wave propulsion (GWP) (Fontana, 2000 and 2003a; Baker, 2000). This is still an untested technique, nevertheless it is based on a straightforward application of the theory of General Relativity (GR), which is trusted by the vast majority of scientists. The new technology is based on High Frequency Gravitational Waves (HFGWs).

A large literature exists on colliding gravitational waves (Szekeres, 1972; Ferrari, 1988a and 1988b), it has been found that the collision or focusing (Alekseev, 1995 and 1996) of gravitational waves produce curvature singularities. These singularities have properties very similar to those of a black-hole, an essential and fundamentally simple object, which produces a gravitational field. GWP is the application of these theories to space travel.

Generators of GWs could be installed directly onboard or remotely to a spacecraft to induce curvature singularities near the spacecraft. The use of HFGW "… as a source of some additional gravitational field…" at a distance was suggested by L. D. Landau and E. M. Lifshitz (1975). According to GR, spacecraft mass interacts with spacetime curvature, therefore the spacecraft will move towards the singularity. In the Newtonian picture, because of the non-linearity of space, the wave at the focus is converted to a Coulomb-like gravitational field. The induced field accelerates the spacecraft and accelerates the reaction mass, this latter can be easily identified with distant masses (Fontana, 2003a). In Newtonian gravity the propagation of the gravitational field is instantaneous; this might not be true, nevertheless it must be accepted when dealing with this approach.

## Generators of Gravitational Waves

To create spacetime singularities, powerful and efficient generators of GW are required. The produced GWs must have a wavelength of the order of millimeters or less, because it is necessary to collimate or focus the waves to a small spot to reach the high energy densities that produces non-linearity effects culminating with the creation of spacetime singularities. The wavelength of GWs must be much smaller than the size of the generator which, in turn, must be compatible with the size of a typical spacecraft.

A large literature describes different proposed devices capable of producing HFGWs (Baker , 2004; Pinto, 1988). Generators of high frequency gravitational waves based on quantum effects may offer higher efficiencies because only at the quantum level there is a precise distinction between gravitational and electromagnetic transitions. In addition, quantum generators may benefit from coherence and/or Bose condensation effects, which may strongly increase the emission probability of GW quanta. Quantum generators of HFGWs can potentially be very efficient in converting input power to GWs.

After recognizing that a quantum source can be developed (Halpern, 1964; Ford, 1982), the first serious difficulty in designing a quantum source of HFGW is the identification of a suitable active material.

The first interesting result consists of recognizing that the orthorombic cuprate High Temperature SuperConductor (HTSC) might become an active material in which gravitational transitions may take place (Fontana, 2000). Later a more comprehensive approach to the problem, including the details of the HTSC GASER, was developed and presented by Fontana and Baker (2003b).

According to the historical nomenclature the quantum source of gravitational waves has been named GASER and the quantum of gravitational wave, graviton.

In the article by Fontana and Baker (2003b), the instanton, with its theoretical foundations developed in Gibbons (1993) (Sanchez, 1982 and 1983), has been used to describe the wave-particle associated to a quantum of gravitational radiation. The (gravitational) instanton can be defined as a pseudoparticle solution of the Einstein equations in Euclidean spacetime (signature ++++) with a cosmological constant. The Euclidean signature helps to remove singularities found in the theory with Lorentzian signature (-+++) and has been introduced for the purpose of the analysis. The Euclidean signature is characterized by imaginary-valued time, the Lorentzian signature is characterized by real-valued time. Research programs have also found instantons solutions in our Lorentzian spacetime and, interestingly, some of them can be identified with colliding gravitational waves (Sanchez, 1982).

The well-known "graviton" is not so well defined. It is usually associated to a spin-2 field that has been somehow quantized. Recently it has been discovered that the graviton may have a mass, depending on boundary conditions like large scale, or local, properties of space-time associated to the $\Lambda$ term, known as the cosmological constant (Novello, 2003), with the result that $m_g^2 = -2\Lambda/3$. The possibly changing sign of the $\Lambda$ term, while going from the vicinity to the source to the free space of our expanding universe, may give unexpected properties to the graviton.

Today, a generally accepted mathematical description of the quantum of GWs does not exist. The lack of experimental access to these objects and their interactions might explain the difficulties in finding an acceptable model. In addition to space propulsion models, a quantum source of GWs is also the resource that must be developed for further progress towards a better understanding of gravitation and cosmology.

## THE HTSC GASER

Different from GW quanta themselves, the mechanisms capable of producing GWs, even from microscopic systems, are better known. The theory describing the emission of gravitational radiation from quantum systems has been developed by Halpern (1964) and Ford (1982). Transitions for which the orbital quantum number L changes by ±2 and for which the total quantum number J changes by 0 or ±2 are gravitational quadrupolar transitions, for which the emission of photons is forbidden and the emission of gravitons is allowed. The selection rules for photons are those for which the orbital quantum number L changes by ±1 and for which the total quantum number J changes by 0 or ±1, etc. For instance, for an atom, gravitational transitions are those between orbitals 3d and one characterized by a lower energy among 3s, 2s and 1s. In typical atomic systems gravitational transitions compete with multiple photon transitions, therefore methods for counteracting photon transitions have to be developed. Detailed analysis comparing the emission of gravitons to the emission of photons has shown that the ratio R of the transition probabilities for matrix elements of equal structures is of the order of $1.6 \times 10^{-36}$ for the proton (Halpern, 1964) and of the order of $4.8 \times 10^{-43}$ for the electron. An intermediate value can be found for the muon.

To increase the efficiency, stimulated emission and coherence can improve the apparently negligible emission probability. The device that employs stimulated emission has been named GASER (Halpern, 1964). We adopted (Fontana, 2003b) the same name for the device that employs wavefunction coherence for increasing the emission probability.

According to Fontana and Baker (2003b), the ideal gaser material should provide:

1. A quantum system with two energy levels characterized by a difference in orbital quantum number of 2, in agreement with Halpern and Laurent selection rules.

2. A quantum system possibly composed of densely packed couples of closely interacting electrons.

3. A quantum system in which the two energy levels are respectively populated by objects with exactly the same wavefunction and the same energy in order to have negligible linewidths, thus permitting efficient stimulated emission.

4. A quantum system in which the two energy levels are populated by quantum objects, for which the two wavefunctions are orthogonal in order to prevent photon transitions and tunneling.

5. A quantum system in which population inversion can be achieved to initiate and sustain gaser action.

A candidate material with the enumerated requirements is a cuprate high temperature superconductor (HTSC) with orthorombic crystal structure, being the two quantum systems *s-wave* ($l$=0) and *d-wave* ($l$=2) Cooper-pair condensates (Kouznetsov, 1997) (Fontana, 2000).

In Fontana and Baker (2003b), it has been shown, with the help of three quantum gravitational theories, that the emission probability for gravitons is proportional to the square of the number of Cooper-pairs involved. Assuming Cooper-pair density of the order of $10^{20}$/cm$^3$ or slightly higher (there are ~$5 \times 10^{21}$/cm$^3$ elementary cells in YBCO) the emission probability looks interesting for a volume of the HTSC greater than few cubic centimeters. In addition to that analysis, the recent discovery that Cooper-pair formation in HTSC involves the *4s* to *3d* electron exchange amplitude in Cu atoms (Mishonov, 2003) that shows the electrons of the pairs posses the quantum numbers required for the gravitational transition. The main process allowing direct transitions between *s-wave* and *d-wave* condensates is therefore purely gravitational.

For direct transitions the energy of the emitted particle only depends on the energy gap. Taking into account the measured *s-wave/d-wave* energy gap, the frequency of 1.3THz can be estimated from BSCCO data (Moessle, 1999).

Like the well-known optical laser, a GASER could be pumped by another GW source, this possibility was not yet properly explored and does not solve the problem of producing GWs from a different form of energy. According to Fontana and Baker (2003b), the HTSC GASER can be pumped electrically, with the injection of *s-wave* Cooper-pairs by a Josephson junction between a Low Temperature SuperConductor (LTSC) and the HTSC, respectively connected to the poles of a current generator. LTSCs are characterized by pure *s-wave* pairing. The estimated power emitted in a GW is of the order of 10 W/ cm$^2$ with a current of 10 kA/cm$^2$.

## FABRICATION OF A PROTOTYPE HTSC GASER

Generally, the fabrication of prototype devices is a complex procedure. It starts with the design of the device, structure and size, goes through the definition of the materials, nature and purity, the definition of the fabrication process, techniques and steps. Finally, the properties of the produced device will be measured to recognize if the results agree with expectation. The procedure is usually repeated many times.

The structure of the proposed HTSC GASER is quite simple; an orthorombic cuprate HTSC mono-crystal should be produced along with a Josephson junction with a LTSC. Both sides of the device should be metalized for connecting the GASER to a current generator. To define the preferable geometry of the HTSC mono-crystal we make some considerations on directivity patterns in wavefunction coherent systems.

Directivity of emission is generally determined by a combination of geometrical properties of the emitter, for instance the shape of the resonant cavity of a LASER. For a wavefunction coherent system in the solid state, composed of microscopic oscillators emitting waves, the directivity function is the product of two components: the directivity function of the single emitter, and the superposition function (interference) of many such emitters. The collective wavefunction is the mechanism that synchronizes the oscillators.

In Halpern (1964) it has been shown that there exists a directivity pattern for transitions emitting gravitons, the directivity here depends on the *d-wave* wavefunction orientation. In HTSC, the macroscopic wavefunction has been observed by interference effects in Josephson junctions. The macroscopic wavefunction is oriented according to the crystallographic orientation of the HTSC crystal. Even if this factor has a minima and maxima, its dependence on the angles is smooth, and it appears to be not critical, nevertheless crystals with different crystallographic orientation should be built to test the theory. The directivity factor related to the superposition function of many emitters is what mainly determines the directivity function and the directivity gain. The angular distribution of GWs for a quantum system is derived in Halpern (1964). For J=M=2 it takes the form:

$$\cos^8\left(\frac{\vartheta}{2}\right) + \sin^8\left(\frac{\vartheta}{2}\right). \tag{1}$$

It has been found that the maximum amount of gravitational radiation occurs in a direction for which the corresponding electromagnetic radiation is excluded or is a minimal. The *d-wave* wavefunction in HTSCs develops in the *a-b* crystallographic axes (Narlikar, 2003), making a maximum of the gravitational radiation emitted in the direction of the *c* axis. This emission is bi-directional.

### Structure

The structure of the proposed HTSC GASER is the one proposed in Fontana and Baker (2003b).

The superposition of a large number of coherent sources is expressed by the interference pattern that they project on a detector at a large distance with respect to the wavelength λ. For instance, the maximum emission for a linear array is for *a* sinϑ=*n* λ, where *a* is the spacing of the point sources, and *n* is an integer to which the maximum is associated. The amplitude of the maxima is N$^2$ times the amplitude produced by each source, where N is the number of sources. More generally a circular planar surface distribution of point sources will produce a beam of radiation orthogonal to the plane of the point sources provided that the wavelength is much shorter than the diameter of the distribution. The emission is bi-directional. A binary neutron star pair is a celestial example in which the orbit plane is the source of bi-directional GW normal to that orbit plane.

Interestingly it is possible to conceive a distribution of coherent point sources on a spherical surface, as shown in FIGURE 1. For outgoing waves they mimic a point source located at the center of the sphere with a power equal to the coherent summation of the radiation emitted by all the point sources. The radiation directed towards the center is focused to the center of the sphere, therefore a large power density can be obtained. The order of magnitude of the power density gain for a given total input power is $A/\lambda^2$, where A is the total area covered by the point sources.

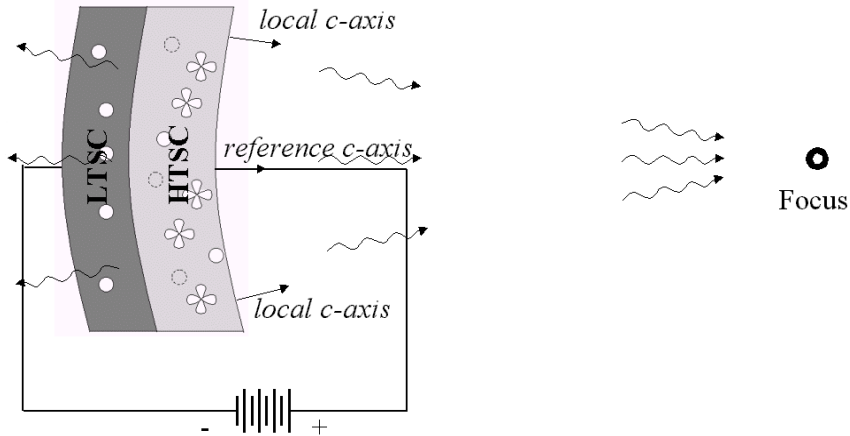

**FIGURE 1**. The Curved HTSC GASER (Section). The Section is Revolved around the *Reference c-axis*.

To produce curved wavefronts it is necessary to produce a thin HTSC film with the required crystallographic orientation, followed by bending the crystal as required to keep the wavefunction coherence on a curved spherical surface. Alternatively a more complex structure should be devised, based on a spherical surface tiled with planar polygons of HTSC interconnected among each other, each with the c-axis directed toward the focus.

The HTSC GASER is a dual layered device, the HTSC *s-wave/d-wave* layer is the one requiring the highest care, because the crystallographic orientation is important for the device to work properly. We therefore need to discuss the HTSC layer with more detail.

## Production techniques and materials.

Orthorombic cuprate HTSC mono-crystals can be produced using different techniques. For bulk material, which appears suitable to planar devices, the melt-textured techniques seem acceptable. This technique requires the sintering of the precursor powders to produce unaligned, grainy HTSC. Afterward, a thermal process termed incongruent melting is performed. The grainy HTSC melts and reacts to form another solid plus a liquid phase. An oriented solidification is then performed, using a $SmBa_2Cu_3O_x$ seed to facilitate the crystallization with the required orientation. Very precise thermal gradients and time-temperature processing are involved. The melt-textured techniques and the related different methods are reviewed by Desgardin (1999).

Thin film techniques seem better suited for making curved GASERs, unfortunately there is no need in current technological applications of curved HTSC films similar to those required here, therefore a new production technology should be developed. Thin film depositions are reviewed by Narlikar (2003). Silicon or Aluminum substrates, buffered by thin layers of $CeO_2$ and Y-stabilized $ZrO_2$ can be coated with HTSCs. Thermal coevaporation may offer 20-30 nm/min growth rates. Pulsed laser deposition (PLD) offers similar rates. Magnetron sputtering is widely used for the deposition of layers of metals with high melting temperatures, it easily adapts to the deposition of HTSC with rates of about 10 nm/min. Metalorganic chemical vapor deposition (MOCVD) may offer 1 µm/min. Sol-gel techniques have also been successfully employed for the preparation of thin films of HTSCs.

Because the coherence gain is proportional to the volume of the HTSC, the melt-textured HTSC seems the best choice for a possible first prototype.

The production of HTSC/LTSC Josephson junction is a standard technique, see for instance the article by Moessle (1999). The Pb LTSC can be evaporated on the HTSC surface directly or with the interposition of an Au or Ag interlayer having thickness between 1 and 6 nm. The HTSC GASER must operate at temperatures below the lowest transition temperature of both HTSC and LTSC, i.e. liquid helium for Pb LTSC. The recently discovered superconductor $MgB_2$ is an extremely interesting option for the LTSC layer because of its high transition temperature of 39 K (Canfield, 2003) and low cost.

The isotopic purity of the raw materials used for sintering the HTSC layer appears a key factor for successfully manufacturing the HTSC GASER. Superconductors in which phonons play a role in inducing electron pairing exhibit a well known isotope effect (Bardeen, 1957; McMillan, 1968). The situation in HTSCs is not completely understood, with experiments giving controversial results, with the latest ones confirming the isotope effect (Bernhard, 2003). The HTSC GASER is based on a very precise transition between two energy levels (macroscopically related to critical temperatures $Tc \propto M^{-\alpha}$, with M mass of the element and $\alpha$ isotope effect). Isotopic mixtures are detrimental at the microscopic scale, where the wavefunction is perturbed with reduction of coherence.

## Experimental setup and driving electronics.

It is possible to conceive a setup based on a planar, disk shaped, GASER with area of about 100 $cm^2$ and thickness of the order of 1 cm. A planar thyristor (SCR – semiconductor controlled rectifier) or a gas thyratron is connected in series with the HTSC GASER and a large set of parallel connected capacitors (C) close the circuit. The capacitors are distributed all around the sandwich. The configuration is intended for pulsed operation with high current. It offers low circuit inductance (L) and low series resistance (R), and is electrically similar to the configuration used for pulsed laser diodes. If the Q of the series equivalent RLC circuit is higher than 1, with a single discharge the GASER will emit a series of exponentially damped pulses of HFGWs with modulation frequency of $1/2\pi\sqrt{(LC)}$. To permit the complete discharge of the capacitors, the thyristor must conduct for the full discharge time and a recirculation diode should be connected in parallel with it. GaAs-based semiconductors can operate at liquid helium temperatures, instead Si-based semiconductors can operate at liquid nitrogen temperatures.

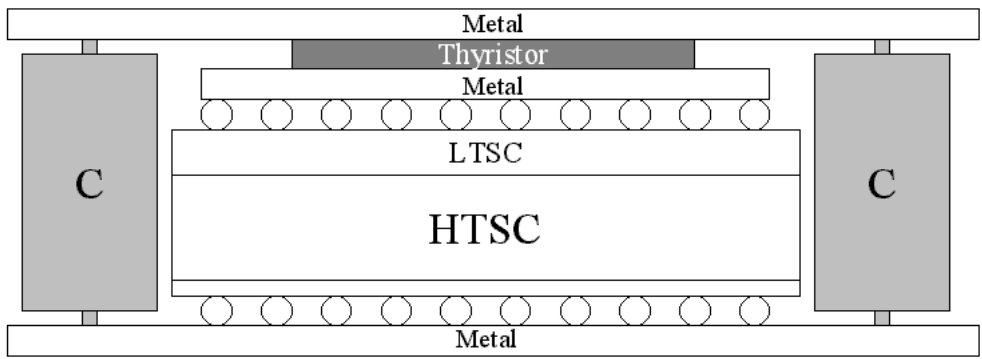

**FIGURE 2.** The Proposed Experimental Setup (Vertical Section).

FIGURE 2 shows a vertical section of the experimental setup for pulsed operations. Differently from standard Josephson junctions used in research instrumentation, the HTSC GASER is a high power device with current densities up to 10 $kA/cm^2$. Indium or Bismuth collapsible ball contacts may facilitate the connection of the device to the GASER cage and driving electronics. A voltage generator should be connected to the top and bottom electrodes for charging the capacitors, it should be disconnected before the discharge. Assuming $Q=R^{-1}\sqrt{(L/C)} \gg 1$, the peak current is $I_P=V_C\sqrt{(C/L)}$. With capacitance of the order of 10 $\mu F$ and inductance of the order of 0.1 $\mu H$, peak currents of the order of 100 kA can be obtained with about 10000 V of charging voltage.

## DETECTION OF HFGWs AND EXPECTED RESULTS.

Gravitational waves, possibly emitted by the HTSC GASER, can be detected by specialized detectors for HFGW in the GHz to THz range. It has been shown that the detection of 1.3 THz HFGW with $h \geq 10^{-30}$, as it is expected from the proposed planar HTSC GASER, could be possible with the synchroresonance converter described by Li (2003). Many different detector are also described in the literature as reported by Baker (2004) and in the book by Misner (1973) and many other articles (Thorne, 1980a-b), unfortunately many detectors have been developed for low frequency GWs and it is unclear if they can be adapted to the THz range. Additionally, it should be emphasized that the "classical" theory of gravitational wave propagation and detection is based on the solution of the linearized Einstein equations, intended to treat small perturbations on flat spacetime. Unfortunately for the approach, our spacetime is not flat but is characterized by a constant curvature ($\Lambda$ term, de Sitter spacetime). Moreover, the energy density near the microscopic sources in the GASER may favor nonlinear effects. Usual simplification conditions seem violated in our approach, which has the potential to make the "classical" detection uncertain.

The expected instanton solutions (Gibbons, 1993; Sanchez, 1982), massive graviton solutions in spacetime with a cosmological constant (Novello, 2003) and self-interaction effects (Ferrari, 1988a and 1988b; Veneziano, 1987; Alekseev, 1996) might indeed simplify the detection of GASER radiation. Standard gravimeters and accelerometers could be the detection instruments. Invoking self-interaction effects, the curved GASER with the parameters already described might produce E$\approx$1mJ of gravitational energy, which is able to accelerate test masses in the vicinity of the focus. The possibility of controlling the operations of the source allows the adoption of lock-in techniques for synchronous detection, thus improving the signal to noise ratio with long integration time.

Claims of observation of gravitational anomalies near HTSC excited by electromagnetic fields have already appeared in the scientific literature (Podkletnov, 1992). Unfortunately, it was not possible to replicate the results till now. This fact might be explained with the incomplete understanding of the phenomenology. Woods (2004) presents a review covering the subject and, more generally, the claims of interaction between gravitation and HTSC.

## CONCLUSION

High Frequency Gravitational Waves have great potential for old, new and unexpected applications. Many interesting and new phenomena have been predicted to become possible by using HFGW. Among them we may cite the possibility to communicate directly through objects opaque to electromagnetic radiation and the possibility of 3D imaging of the internal structure of large and very dense objects like the Earth itself (Baker, 2000). Space travel is an old application that may strongly benefit from HFGW. As soon as powerful generators of HFGW will be made available, spacecrafts capable of fast interstellar travel could be constructed. It is superfluous to cite the extraordinary benefits of the discovery of extra-solar planets and their possible life forms. Commercial airlines could operate a fleet of planes with HFGW propulsion, permitting a faster and more comfortable trip to the destination. The HTSC GASER is the proposed device capable of generating ~1 THz HFGWs with high conversion efficiency and is intended to start the new era. This article has presented an approach for the construction of a proof-of-concept prototype, comprising the design of the GASER structure, choice of the materials and basic electronics designed for pulsed operations. Techniques for the detection of the emitted radiation have also been discussed.

## ACKNOWLEDGMENTS

The author whishes to acknowledge Dr. Franklyn Mead and AFRL for a grant to attend the STAIF 2004 Conference.